\documentclass[11pt]{article}
\usepackage{xspace}
\usepackage{graphicx}
\usepackage{amsmath}
\usepackage{amssymb}
\usepackage{color}
\usepackage{hyperref}
\usepackage{bookmark}

\definecolor{Red}{rgb}{1,0,0}
\definecolor{Green}{rgb}{0,1,0}
\definecolor{Blue}{rgb}{0,0,1}
\definecolor{Black}{rgb}{0,0,0}

\def\blue{\color{Blue}}

\def\beq{\begin{equation}}
\def\eeq#1{\label{#1}\end{equation}}
\def\eeqn{\end{equation}}

\def\beqa{\begin{eqnarray}}
\def\eeqa#1{\label{#1}\end{eqnarray}}
\def\eeqan{\end{eqnarray}}

\def\leqn#1{(\ref{#1})}

\let\bar=\overbar

\def\O{{\cal O}}

\def\Dslash{\not{\hbox{\kern-4pt $D$}}}
\def\dslash{\not{\hbox{\kern-2pt $\del$}}}

\def\msb{{\bar{\ssstyle M \kern -1pt S}}}

\newcommand\vud {\ensuremath{V_{ud}}\xspace}
\newcommand\vus {\ensuremath{V_{us}}\xspace}
\newcommand\vub {\ensuremath{V_{ub}}\xspace}
\newcommand\vcd {\ensuremath{V_{cd}}\xspace}
\newcommand\vcs {\ensuremath{V_{cs}}\xspace}
\newcommand\vcb {\ensuremath{V_{cb}}\xspace}
\newcommand\vtd {\ensuremath{V_{td}}\xspace}
\newcommand\vts {\ensuremath{V_{ts}}\xspace}
\newcommand\vtb {\ensuremath{V_{tb}}\xspace}

\def\vckm       {\ensuremath{{V}_{\rm CKM}}\xspace}

\def\Vud  {\ensuremath{|\vud|}\xspace}
\def\Vcd  {\ensuremath{|\vcd|}\xspace}
\def\Vtd  {\ensuremath{|\vtd|}\xspace}
\def\Vus  {\ensuremath{|\vus|}\xspace}
\def\Vcs  {\ensuremath{|\vcs|}\xspace}
\def\Vts  {\ensuremath{|\vts|}\xspace}
\def\Vub  {\ensuremath{|\vub|}\xspace}
\def\Vcb  {\ensuremath{|\vcb|}\xspace}
\def\Vtb  {\ensuremath{|\vtb|}\xspace}

\newcommand{\ph}[1]{\ensuremath{\phantom{#1}}}
\newcommand{\CP}{\ensuremath{C\!P}\xspace}
\makeatletter
\newcommand*{\rom}[1]{\expandafter\@slowromancap\romannumeral #1@}
\makeatother
\newcommand{\Bs}{\ensuremath{B^0_s}\xspace}
\newcommand{\Bsb}{\ensuremath{\bar{B}^0_s}\xspace}
\newcommand{\Dsm}{\ensuremath{D^-_s}\xspace}
\newcommand{\Dsp}{\ensuremath{D^+_s}\xspace}
\newcommand{\Dz}{\ensuremath{D^0}\xspace}
\newcommand{\Dzb}{\ensuremath{\bar{D}^0}\xspace}
\newcommand{\Ks}{\ensuremath{K^0_S}\xspace}
\newcommand{\BtoDh}{\ensuremath{B^\pm \to Dh^\pm}\xspace}

\newcommand{\BstoDsK}{\ensuremath{B^0_s \to D^\mp_sK^\pm}\xspace}

\def\Title#1{\begin{center} {\Large {\bf #1} } \end{center}}

\begin{document}

\Title{Measurement of the CKM angle \(\gamma\) at LHCb}

\bigskip\bigskip

\addtocontents{toc}{{\it S. Ali}}
\label{AliStart}

\begin{raggedright}  

%% Authors - you should specify at least one author as follows.
Suvayu Ali\index{Ali, S.}, {\it FOM-Nikhef}\\

%% In case you want to quote your collaboration please modify the text below.
%% If this is not relevant then you may comment out the following line.
\begin{center}\emph{On the behalf of the LHCb Collaboration.}\end{center}
\bigskip
\end{raggedright}

{\small
\begin{flushleft}
\emph{To appear in the proceedings of the 50 years of CP violation conference, 10 -- 11 July, 2014, held at Queen Mary University of London, UK.}
\end{flushleft}
}

%% begining of text

\section{The CKM angle \texorpdfstring{\(\gamma\)}{gamma}}
\label{sec-1}
The Standard Model (SM) describes all known fundamental particles and
their interactions, excluding gravity.  It has survived rigorous
experimentation over four decades.  However some observations, like
baryon matter asymmetry in the universe, remain inadequately
explained.  This asymmetry requires breaking of Charge-Parity (\CP)
symmetry \cite{Sakharov:1967dj}, among other requirements.

In 1973, M. Kobayashi, and T. Maskawa \cite{Kobayashi:1973fv},
proposed the Kobayashi-Maskawa mechanism, now known as the
Cabibbo-Kobayashi-Maskawa (CKM) formalism.  In this formalism, the CKM
matrix quantifies the couplings between quarks of different flavour.
It is a transformation between quark mass and flavour eigenstates:
\begin{equation}
  \label{eq:eigenstates}
  \begin{bmatrix}
    d' \\ s' \\ b' \\
  \end{bmatrix}
  = \underbrace{
    \begin{bmatrix}
      V_{ud} & V_{us} & V_{ub} \\
      V_{cd} & V_{cs} & V_{cb} \\
      V_{td} & V_{ts} & V_{tb}
    \end{bmatrix}
  }_{\vckm}
  \begin{bmatrix}
    d \\ s \\ b \\
  \end{bmatrix}.
\end{equation}

Presence of complex phases in some of the couplings lead to \CP
violation in particle decays.  Expressing the CKM matrix using the
Wolfenstein parametrisation \cite{Wolfenstein:1983yz},
\begin{align}
  \label{eq:Vckm}
  \vckm &=
  \begin{bmatrix}
    \ph{-}\Vud\ph{e^{-\iota\beta}} & \ph{-}\Vus\ph{e^{\iota\beta_s}} & {\blue\Vub e^{-\iota\gamma}} \\
    -\Vcd\ph{e^{-\iota\beta}}      & \ph{-}\Vcs\ph{e^{\iota\beta_s}} & \Vcb\ph{e^{-\iota\gamma}} \\
    \ph{-}\Vtd e^{-\iota\beta}     & -\Vts e^{\iota\beta_s}          & \Vtb\ph{e^{-\iota\gamma}}
  \end{bmatrix}+ \O(\lambda^5),
\end{align}
provides a description of the \CP violating particle decays involving
the complex phases on \vub, \vtd, and \vts.

Imposing unitarity conditions allows us to represent the CKM matrix by
six triangles on the complex plane.  Precise determination of the CKM
triangle is needed to scrutinise the consistency of the SM.  Although
there have been many efforts to measure the triangle before, the angle
\(\gamma\),
\begin{equation}
  \label{eq:gamma}
  \gamma = arg(-\vud\vub^{*}/\vcd\vcb^{*}),
\end{equation}
remains poorly determined.

\section{The LHCb detector}
\label{sec-2}
All analyses discussed here are performed with data recorded with the
LHCb detector at CERN.  The LHCb detector operates in the busy
hadronic environment of the Large Hadron Collider.  The design and
performance of the detector are discussed in greater detail elsewhere
\cite{Alves:2008zz:Aaij:2014pwa}.  It operates as a precise momentum
spectrometer with a pseudo-rapidity coverage in the forward region,
\(2 < \eta < 5\), which covers 40\% of the beauty production
cross-section.  It offers excellent track reconstruction and decay
vertex resolution making it ideal for studying long-lived particles
like \emph{B}-mesons.  LHCb includes two Ring-Imaging Cherenkov (RICH)
detectors for particle identification, allowing for powerful
discrimination between kaons, pions, and protons.  Together with the
calorimeters and the muon detector, LHCb provides identification of
muons, electrons, and photons.  These capabilities are essential to
study the fully hadronic decay modes described later.

\section{Measuring \texorpdfstring{\(\gamma\)}{gamma} using hadronic \emph{tree} decays}
\label{sec-3}
Past \(\gamma\) measurements were done by the \textsc{BaBar} and Belle
collaborations using coherent production of charged \emph{B}-meson decays.
In 2012, the Belle collaboration reported a \(\gamma\) value of
\((68^{+15}_{-14})^{\circ}\) \cite{Trabelsi:2013uj}.  Similarly, the
\textsc{BaBar} collaboration reported \(\gamma\) to be
\((69^{+17}_{-16})^{\circ}\) \cite{Lees:2013nha}.

In this article I will summarise \(\gamma\) measurements at LHCb from tree
decays of \emph{B}-mesons, as opposed to the determination through
charmless \emph{B}-decays with higher order diagrams \cite{Aaij:2014xba},
leading to the world's most precise determination.  The measurements
can be categorised into two types: decay-time integrated, and
decay-time dependent.  To gain precision, we statistically combine the
different results.  I conclude with an outlook on the future of \(\gamma\)
measurements.

\subsection{\texorpdfstring{\(\CP\)}{CP} violation in decay-time integrated \texorpdfstring{\(\BtoDh\)}{B->Dh}}
\label{sec-3-1}
There are several well established methods to determine \(\gamma\) from
\BtoDh decays, where \emph{D} stands for an admixture of \Dz and \Dzb.  The
methods use asymmetries in decay-time integrated decay amplitudes as one
of the observables of interest.  I present a selection of such
analyses below.

\begin{figure}[!htb]
\centering
\includegraphics[width=0.8\linewidth]{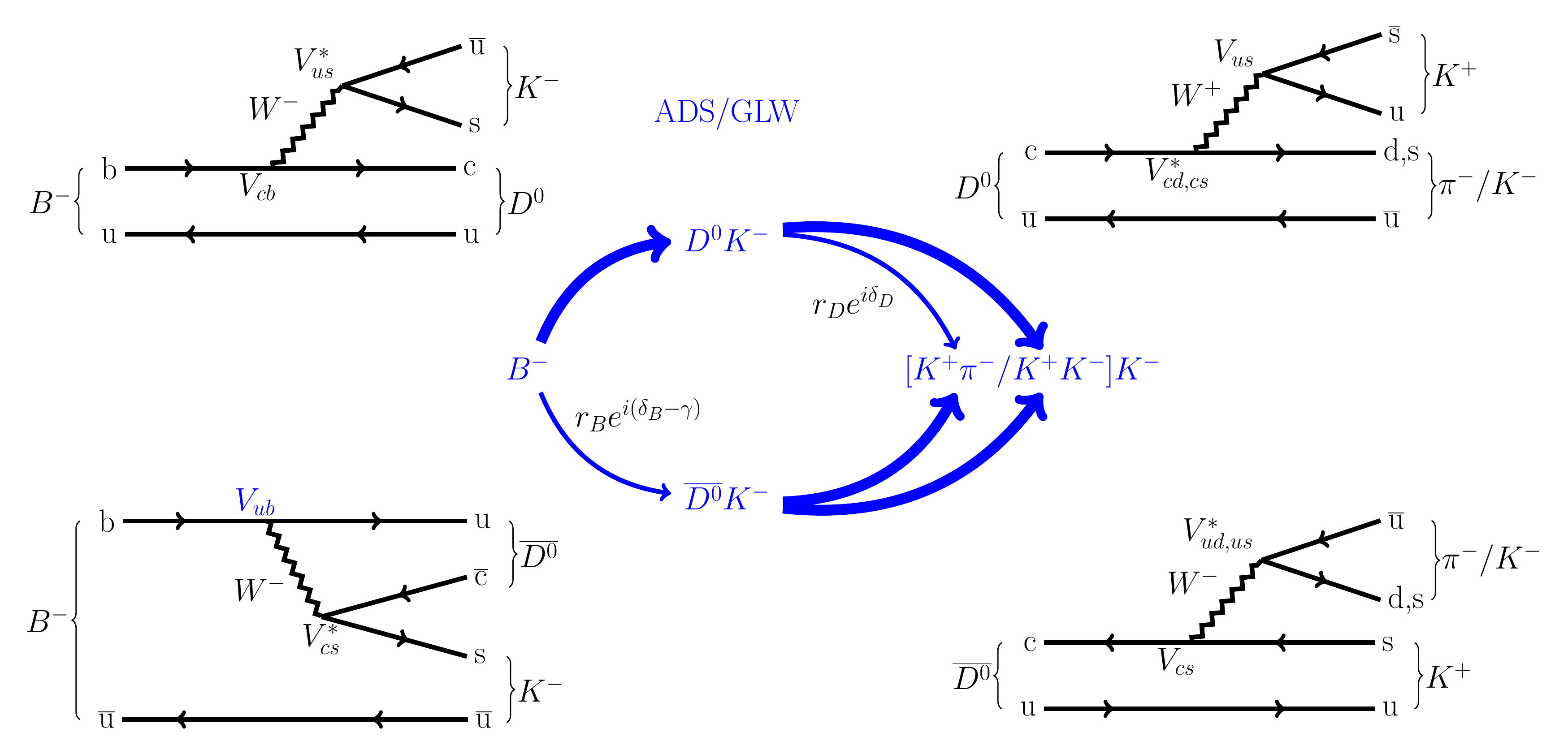}
\caption{\label{fig:ads-glw-schema}Schematic diagram showing relative amplitude and interference of \BtoDh decays considered by the ADS, and GLW methods.}
\end{figure}

\BtoDh decays can provide powerful methods for \(\gamma\) determination.
The amplitude of \(B^- \to \Dz K^-\) is proportional to \vcb, whereas
\(B^- \to \Dzb K^-\) is proportional to \vub (see
Fig. \ref{fig:ads-glw-schema}).  If the \emph{D} final state is accessible
to both \Dz and \Dzb, the two decay paths can interfere and one can
extract observables sensitive to \(\gamma\).

So far only \emph{D} decays where they decay into \CP-even eigenstates
(e.g. \(D \to K^+K^-, \pi^+\pi^-\))
\cite{Gronau:1990ra:Gronau:1991dp}, or other modes like \(D \to
\pi^-K^+\) \cite{Atwood:1996ci:Atwood:2000ck} have been considered.
These two methods are named after the initials of the proponents as
``GLW'', and ``ADS'' respectively.  For the ADS modes, the \(B^- \to \Dz
K^-\) decay is followed by a doubly Cabibbo-suppressed \emph{D} decay,
whereas the suppressed \(B^- \to \Dzb K^-\) is followed by a favoured
\emph{D} decay mode.  This results in comparable amplitudes for both decay
paths, leading to larger interference in comparison to GLW modes.  A
schematic diagram illustrating this effect is shown in
Fig. \ref{fig:ads-glw-schema}.

Notable observables for both of these methods are partial widths
ratios and \CP asymmetries shown below.
\begin{align}
  \label{eq:ads-glw-1}
  R_{\CP+} &= 1 + r_{B}^2 + 2r_{B}\cos\delta_{B}\cos\gamma \\
  \label{eq:ads-glw-2}
  A_{\CP+} &= 2r_{B}\sin\delta_{B}\sin\gamma \;/\; R_{CP+} \\
  \label{eq:ads-glw-3}
  R_{ADS}  &= r_{B}^2 + r_{D}^2 + 2r_{B}r_{D}\cos(\delta_{B}+\delta_{D})\cos\gamma \\
  \label{eq:ads-glw-4}
  A_{ADS}  &= 2r_{B}r_{D}\sin(\delta_{B}+\delta_{D})\sin\gamma \;/\; R_{ADS}
\end{align}
The variables labelled \(\CP+\) correspond to modes where the \emph{D}
decays to \CP-even eigenstates (GLW), and the variables corresponding
to the ADS method are labelled \emph{ADS}.  \emph{B} decay rate ratios are
called \emph{R}, whereas \CP asymmetries are called \emph{A}.  The observables
in Eqs.  (\ref{eq:ads-glw-1}--\ref{eq:ads-glw-4}) have been expressed
in terms of amplitude ratios (\(r_B\) and \(r_D\)), strong phase
differences (\(\delta_B\) and \(\delta_D\)), and the CKM angle \(\gamma\)
(see Fig. \ref{fig:ads-glw-schema} for an illustration depicting the
role of the physics parameters).

\begin{figure}[!htb]
\centering
\includegraphics[width=0.8\linewidth]{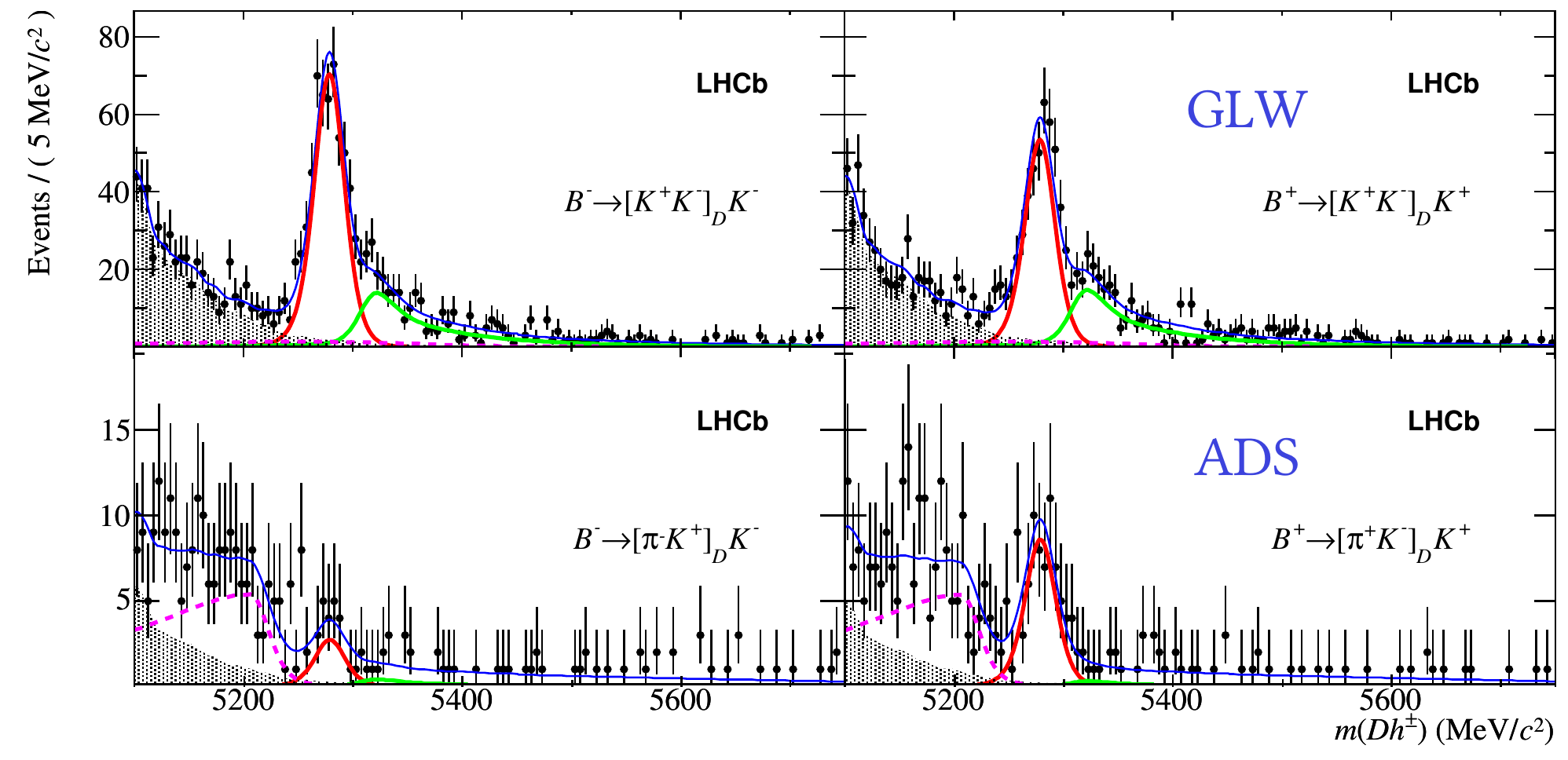}
\caption{\label{fig:ads-glw-plot}\BtoDh events measured using the GLW, and ADS methods.}
\end{figure}

The GLW method has the advantage of having larger event statistics,
owing to favoured \emph{D} decay modes, whereas the ADS method can boast
large intereference due to comparable decay amplitudes.  The invariant
mass distributions from these measurements are shown in
Fig. \ref{fig:ads-glw-plot}.  The top row shows events measured using
the GLW method, and events measured with the ADS method are shown
below.  Analysis of the \(B \to DK\) ADS mode shows evidence of a
large negative asymmetry at 4.0\(\sigma\) significance.  Similarly the
combined \(B \to Dh\) GLW modes show positive asymmetries (4.5\(\sigma\))
\cite{Aaij:2012kz}.  Subsequently, measurements for suppressed ADS
modes, where the \emph{D} undergoes a 4-body decay to \(K\pi\pi\pi\) were
also performed successfully \cite{Aaij:2013mba}.

\begin{figure}[!htb]
  \centering
  \includegraphics[width=0.3\linewidth]{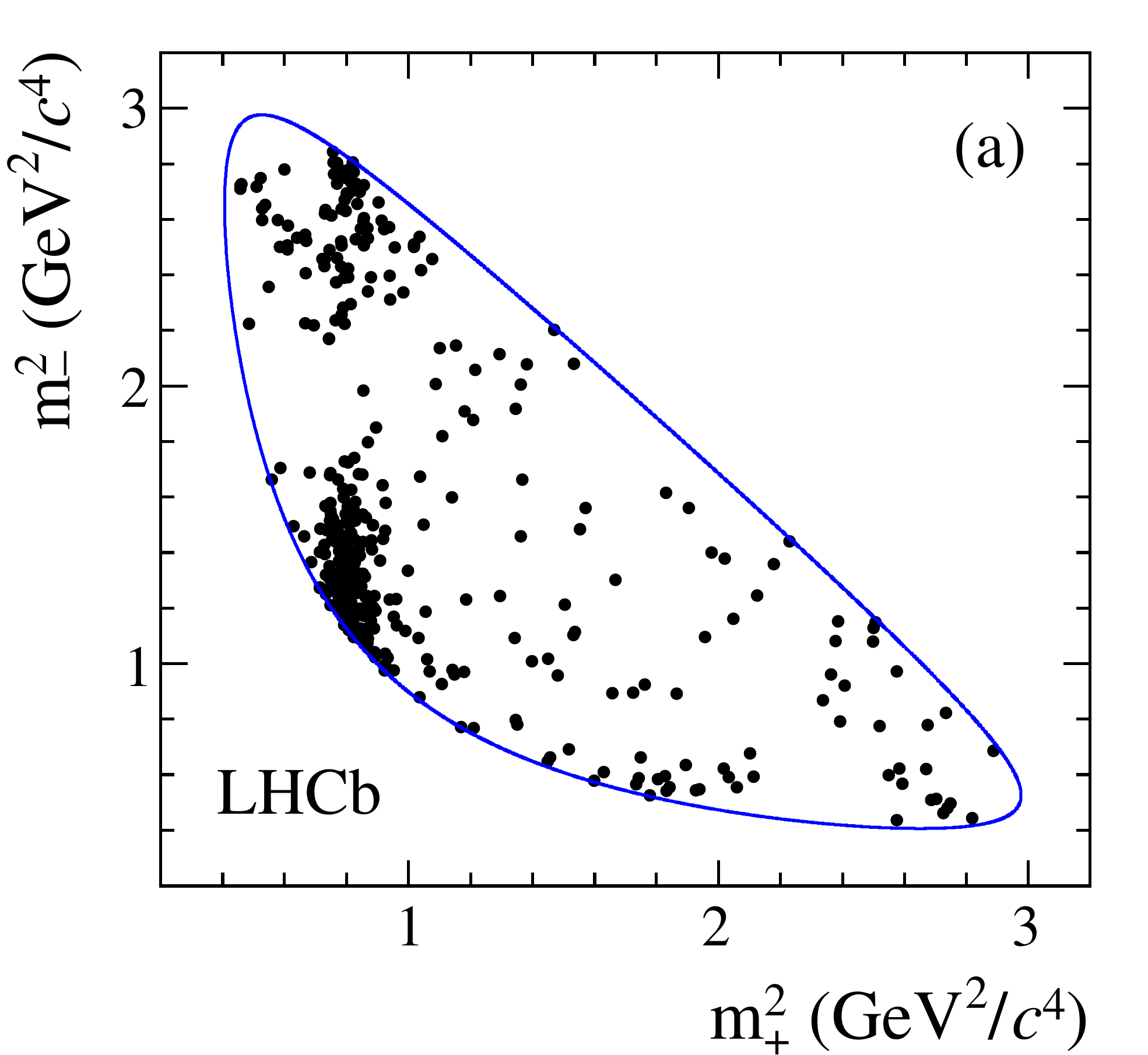}
  \includegraphics[width=0.3\linewidth]{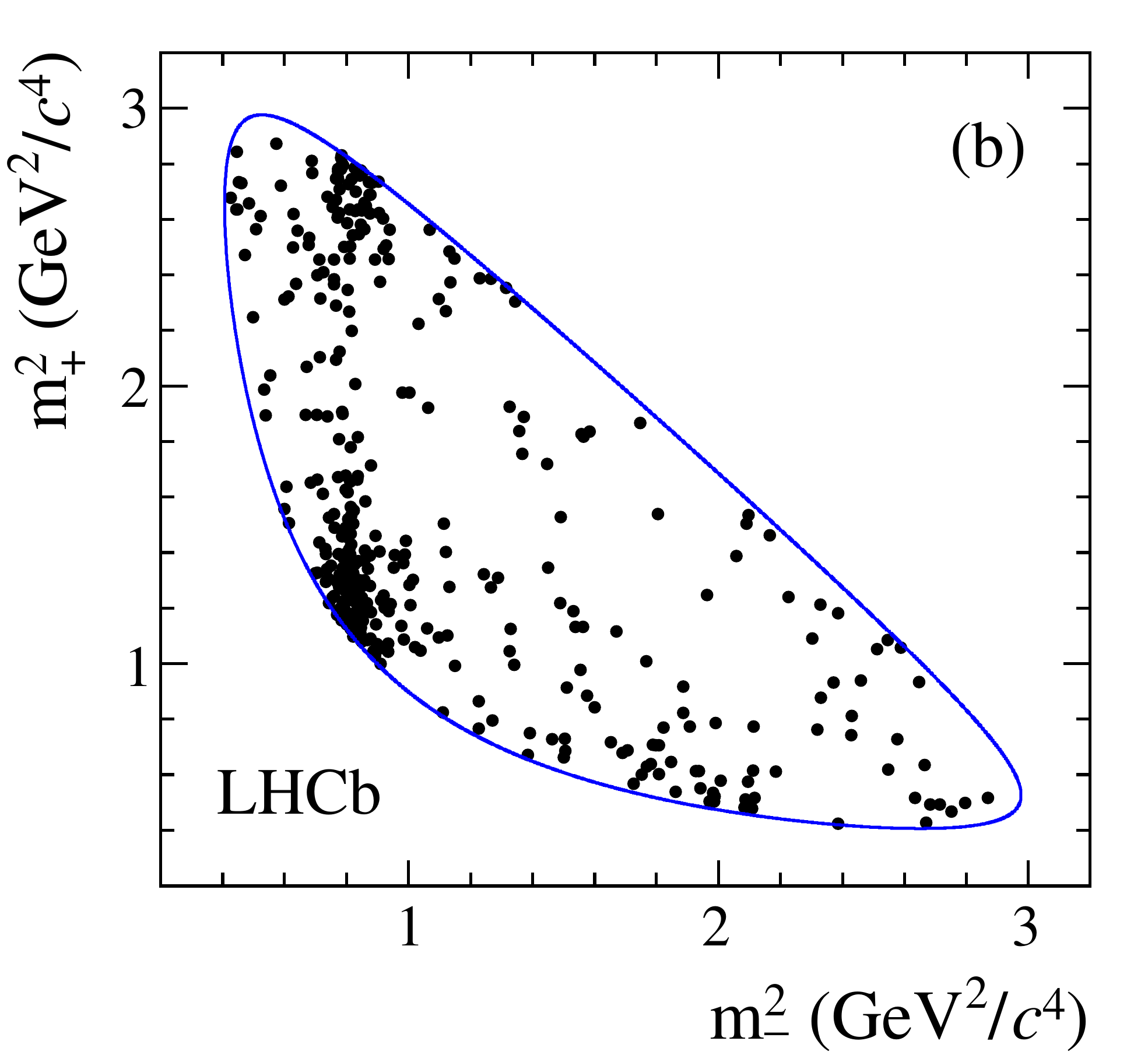}
  \includegraphics[width=0.3\linewidth]{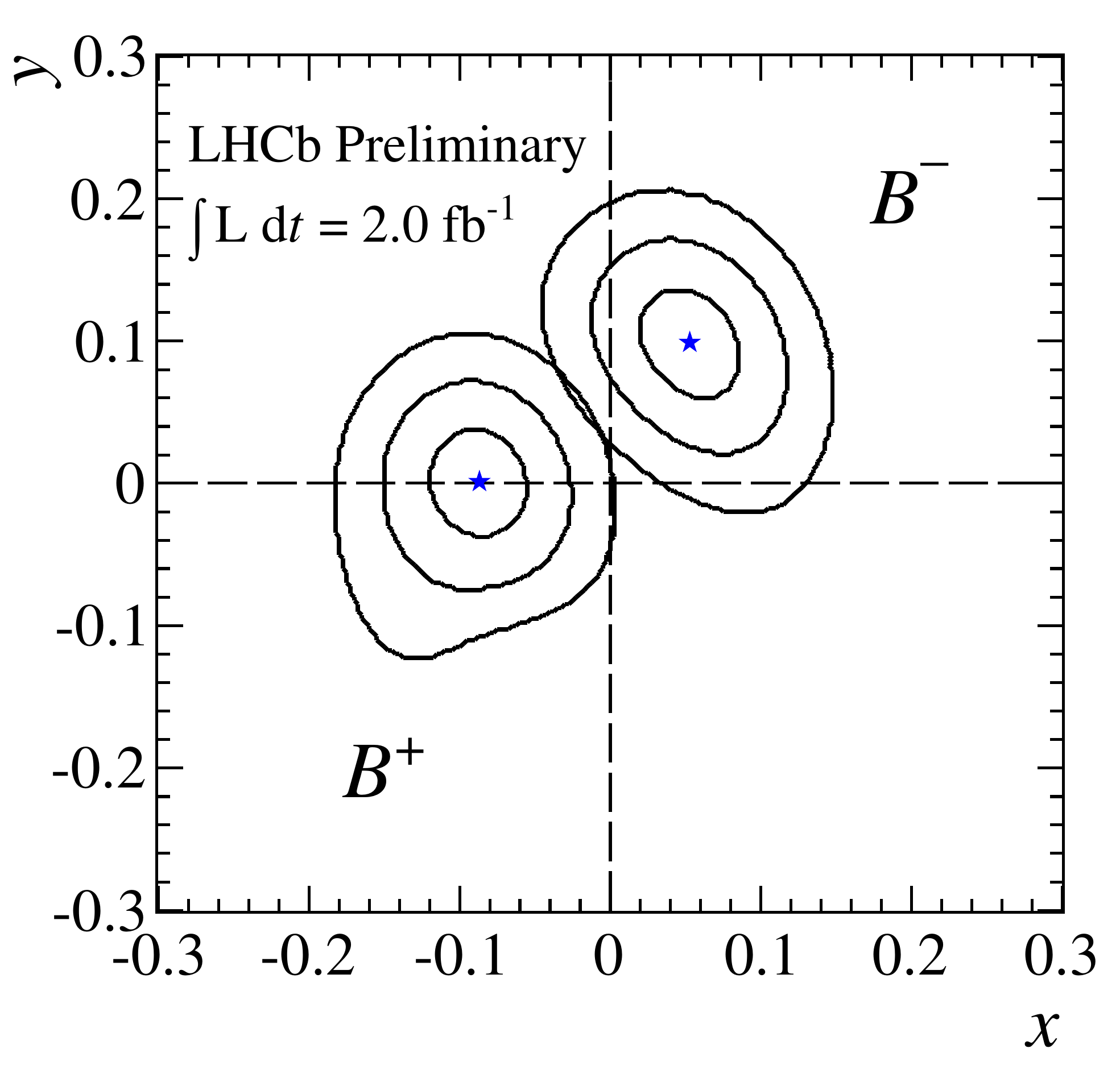}
  \caption{Dalitz plots for \(D \to \Ks\pi^+\pi^-\) separated between
    \(B^+\) (left) and \(B^-\) (middle) decays are shown above.  On
    the Dalitz plane, \(m_+\) stands for the invariant mass
    constructed from the \Ks and the \(\pi^+\), and \(m_-\) is
    constructed from the \Ks and \(\pi^-\).  Best fit values of
    \((x_{\pm},y_{\pm})\) from the GGSZ analysis are shown on the
    right.  Central values are indicated by a star; 1\(\sigma\),
    2\(\sigma\), and 3\(\sigma\) confidence levels are also shown.}
  \label{fig:ggsz-plot}
\end{figure}

Yet another method to extract \(\gamma\) from \(B^\pm \to DK^\pm\) decays,
involves the \emph{D} decaying to a self \CP-conjugate state like: \(\Ks
K^+K^-\) or \(\Ks\pi^+\pi^-\) \cite{Giri:2003ty:BONDARGGSZ},
henceforth collectively referred to as \(\Ks h^+h^-\).  This method is
labelled ``GGSZ'' after the initials of the proponents.  The idea is to
compare \(D \to \Ks h^+h^-\) final states in the Dalitz plane between
\(B^+ \to DK^+\) and \(B^- \to DK^-\) decays.  This method requires a
good understanding of the variation of strong phase in the Dalitz
plane.  Which is known from direct measurements of the decay of
\Dz-\Dzb entangled pairs from \(\psi(3770)\) decays
\cite{Libby:2010nu}, performed by the CLEO-c collaboration.  Fitting
the Dalitz bin contents, we measure the observables:
\begin{equation}
  \label{eq:ggsz}
  x_{\pm} = r_{B}\cos(\delta_{B}\pm\gamma), \quad \text{and} \quad y_{\pm} = r_{B}\sin(\delta_{B}\pm\gamma),
\end{equation}
where \(x_{\pm}\) and \(y_{\pm}\) are Cartesian parameters sensitive
to \(\gamma\).  Analysing the complete 3fb\(^{\text{-1}}\) dataset, we find best fit
values for the observables in Eq. \leqn{eq:ggsz} are consistent with
non-zero.  Fig. \ref{fig:ggsz-plot} shows the best fit values and two
of the corresponding Dalitz plots \cite{Aaij:2012hu}.

\subsection{\texorpdfstring{\(\CP\)}{CP} violation in time-dependent \texorpdfstring{\(\BstoDsK\)}{Bs->DsK}}
\label{sec-3-2}
\begin{figure}[!htb]
\centering
\includegraphics[width=0.8\linewidth]{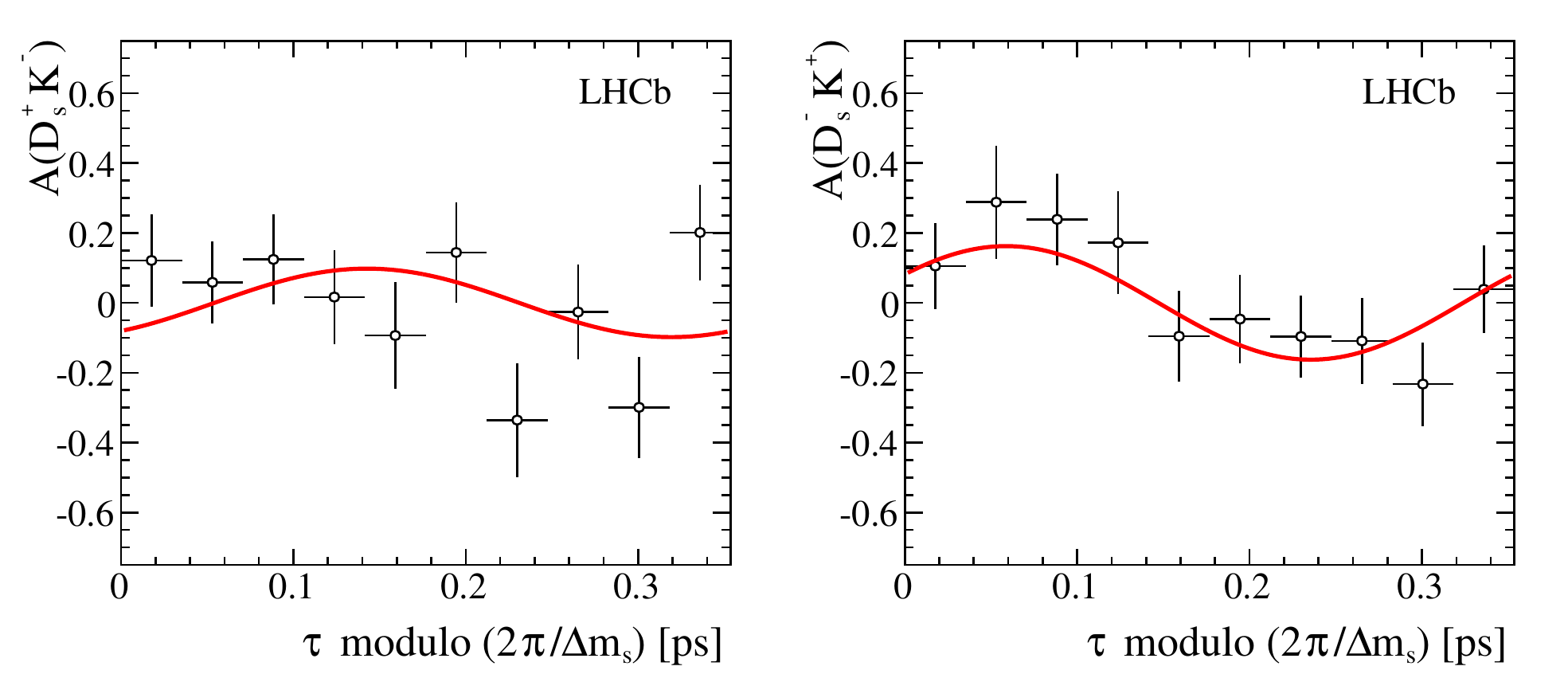}
\caption{\label{fig:dsk-asym}Time-dependent asymmetry for the two final states where the decay-time axis has been remapped to one oscillation period.}
\end{figure}

\BstoDsK decays present an opportunity to measure \(\gamma\) from tree
decays by studying the decay-time distribution of the \(B_s\)-meson
\cite{DeBruyn:2012jp}.  Both \Bs and \Bsb can decay to the two final
states: \(\Dsm K^+\) and \(\Dsp K^-\).  This leads to a superposition
of four decay equations, with five \CP violation parameters which
depend on \(\gamma\).  In this analysis, we first perform a
multi-dimensional mass fit to the \(B_s\)-mass, \(D_s\)-mass, and Kaon
particle identification distributions to determine the proportions of
different physics contribution in our signal mass window.
Subsequently we fit the decay-time distribution and extract the \CP
observables and find \(\gamma = (115^{+28}_{-43})^\circ\),
\(\delta_{D_sK} = (3^{+19}_{-20})^\circ\), and \(r_{D_sK} =
0.53^{+0.17}_{-0.16}\) \cite{Aaij:2014fba}.  Time-dependent asymmetry
plots for the two final states are shown in Fig. \ref{fig:dsk-asym}.

\section{Combination of \texorpdfstring{\(\gamma\)}{gamma} measurements}
\label{sec-4}
We perform a \(\chi^2\) combination of all the experimental inputs
from the \BtoDh decays (including \(B \to D\pi\)).  Results from the
ADS and GLW analyses with 1fb\(^{\text{-1}}\) of data, and the GGSZ analysis
using 3fb\(^{\text{-1}}\) of data are combined.  Effects of \Dz mixing
\cite{Aaij:2012nva} are taken into account and hadronic parameters for
the \Dz from the CLEO collaboration \cite{Lowery:2009id} are used.  We
find \(\gamma = (67 \pm 12)^\circ\) at 68\% C.L
\cite{Aaij:2013zfa:LHCb:2013gka}.

\begin{figure}[!ht]
  \centering
  \includegraphics[width=0.45\linewidth]{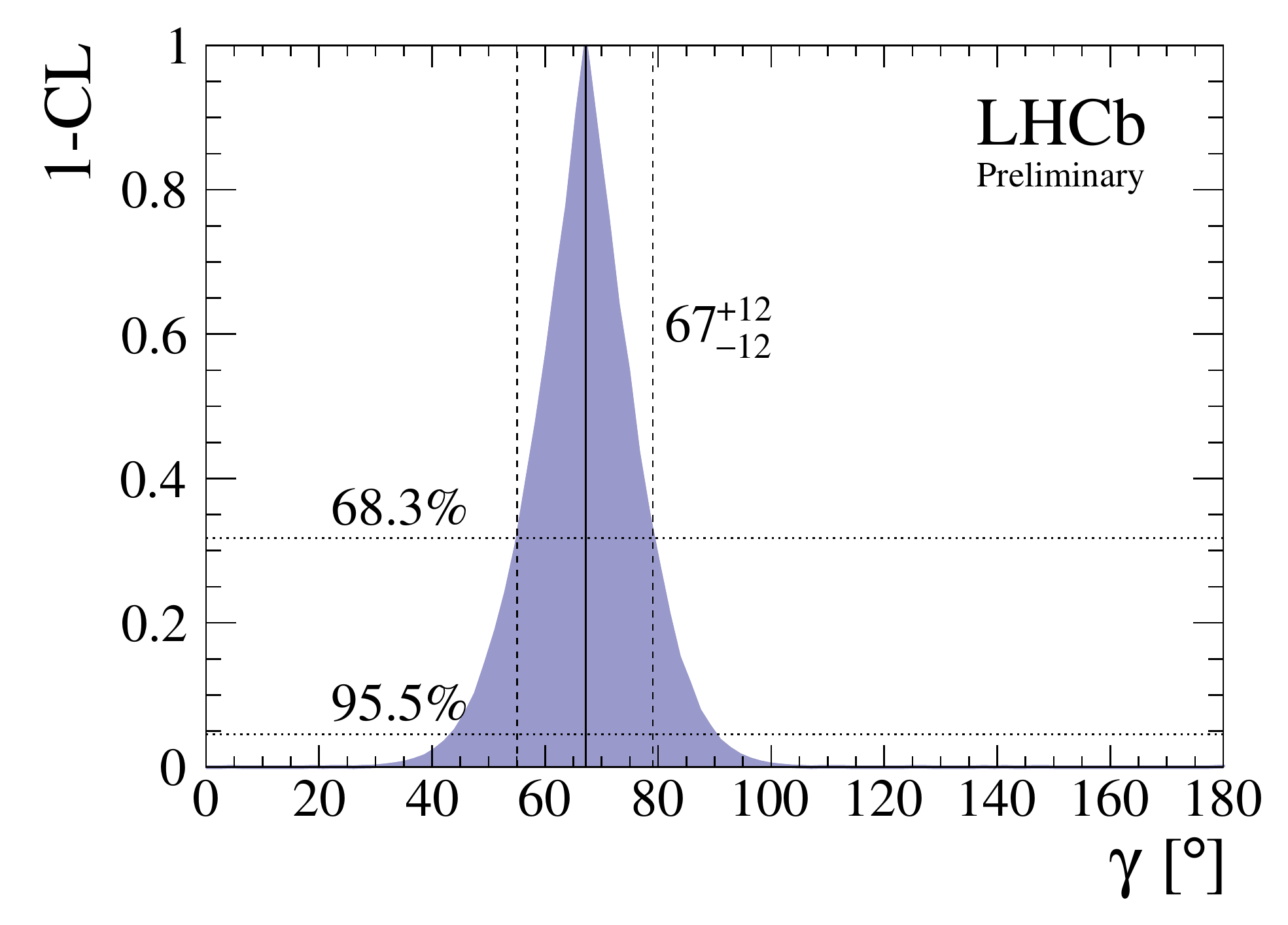}
  \includegraphics[width=0.45\linewidth]{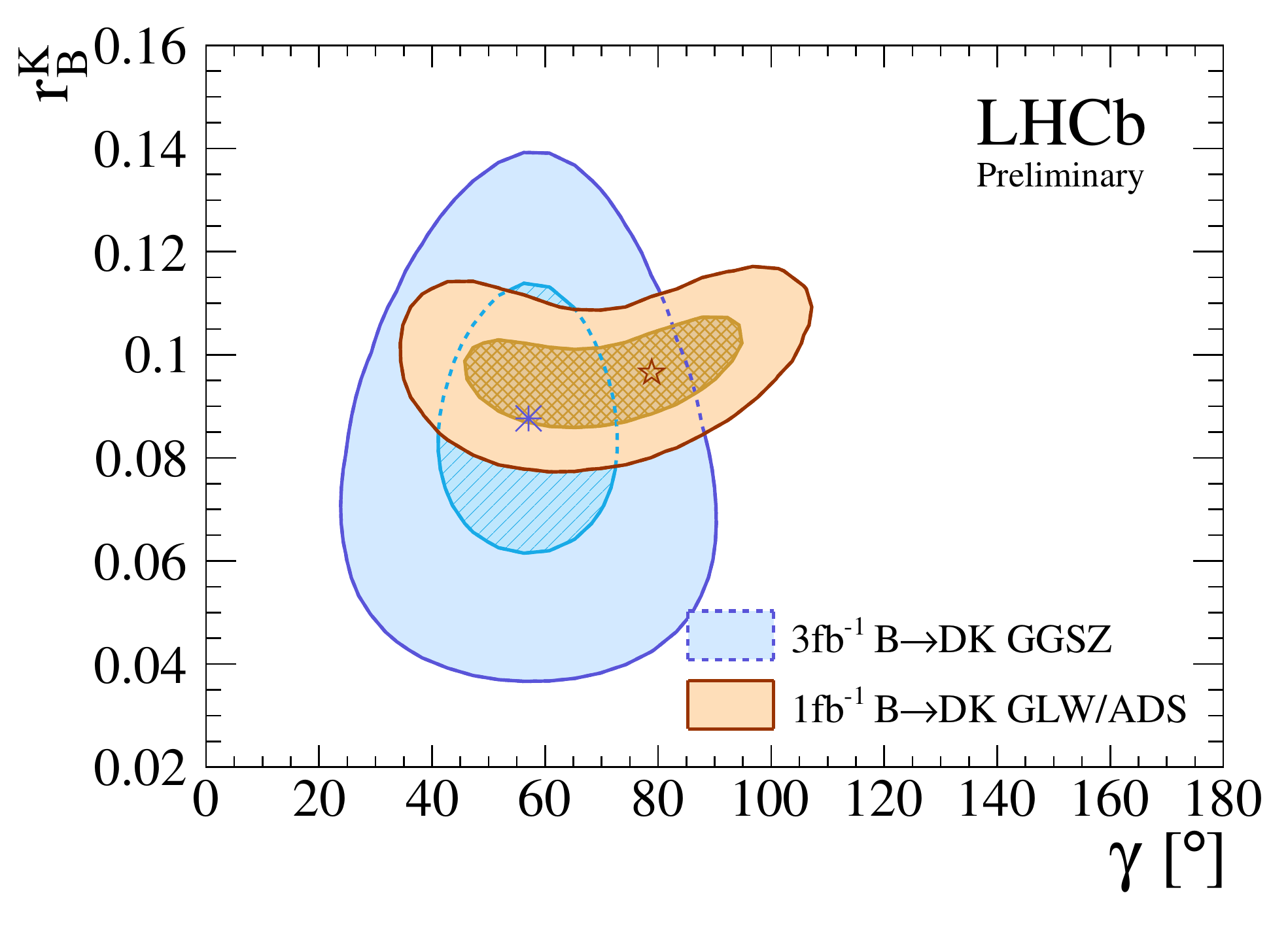}
  \caption{Left: 1-CL distribution for \(\gamma\) for \(B^+ \to
    DK^\pm\) decays.  Best fit values are shown with uncertainties
    corresponding to 68.3\% CL intervals.  Right: Profile likelihood
    contour plot for \(\gamma\).  Contours corresponding to
    1fb\textsuperscript{-1} (orange) and 3fb\textsuperscript{-1}
    (blue) data are shown separately.}
  \label{fig:gamma}
\end{figure}

The result presented above was the world's most precise measurement of
the angle \(\gamma\) at the time of the conference.  However, since then
an updated result of \(\gamma = (73^{+9}_{-10})^\circ\) was presented
by the LHCb collaboration \cite{LHCB-CONF-2014-004}.  The new result
includes other \emph{B} decay modes, such as, \(B \to DK^*\)
\cite{Aaij:2014eha}, and new \emph{D} decay modes for existing channels,
like \(B \to Dh, D \to K^0_SK\pi\) \cite{Aaij:2014dia}.  These two
analyses along with the updated GGSZ analysis \cite{Aaij:2014uva} use
the complete 3fb\(^{\text{-1}}\) dataset.  The updated combination also includes
the decay time-dependent measurement of \BstoDsK decays.  In the long
run, precision of \(\gamma\) measurement at LHCb is expected to improve
significantly.  Table \ref{tab:sensitivity} provides a brief summary.

\begin{table}[htb]
\caption{\label{tab:sensitivity}Expected sensitivity for \(\gamma\) measurements at LHCb from charged B decays, and time-dependent measurements \cite{Bediaga:1443882}.}
\centering
\begin{tabular}{lrr}
 & Run \rom{2} & Upgrade\\
\hline
\(\gamma(B \to DK)\) & \(4^{\circ}\) & \(0.9^{\circ}\)\\
\(\gamma(B_s \to D_sK)\) & \(11^{\circ}\) & \(2^{\circ}\)\\
\end{tabular}
\end{table}

Recently results from a previously unexplored analysis, \(\Lambda^0_b
\to \Dz ph^-\) \cite{Aaij:2013pka}, with a promising possibility to
contribute to \(\gamma\) measurement \cite{Giri:2001ju} was also reported
by LHCb.  To conclude, precision of the measurement of angle \(\gamma\)
has been improving steadily.  LHCb is expected to make significant
contribution towards this goal over the coming years.

%% end of text

\end{document}